# Detecting photoelectrons from spontaneously formed excitons

Keisuke Fukutani[1,2], Roland Stania[1,2], Chang Il Kwon[1,3],

Jun Sung Kim[1,3], Ki Jeong Kong[1], Jaeyoung Kim[1,2], Han Woong Yeom[1,3]*

[1]*Center for Artificial Low Dimensional Electronic Systems, Institute for Basic Science (IBS), Pohang 37673, Republic of Korea*

[2]*Pohang Accelerator Laboratory, Pohang University of Science and Technology (POSTECH), Pohang 37673 Republic of Korea*

[3]*Department of Physics, Pohang University of Science and Technology (POSTECH), Pohang 37673 Republic of Korea*

*Corresponding author. Email: yeom@postech.ac.kr (H.W.Y.)

**Excitons, quasiparticles of electrons and holes bound by Coulombic attraction, are created transiently by light and play an important role in optoelectronics, photovoltaics and photosynthesis. While they are also predicted to form spontaneously in a small gap semiconductor or a semimetal, leading to a Bose-Einstein condensate at low temperature, their material realization has been elusive without any direct evidence. Here we detect the direct photoemission signal from spontaneously formed excitons in a debated excitonic insulator candidate $Ta_2NiSe_5$. Our symmetry-selective angle-resolved photoemission spectroscopy reveals a characteristic excitonic feature above the transition temperature, which provides detailed properties of excitons such as anisotropic Bohr radius. The present result evidences so called preformed excitons and guarantees the exciton insulator nature of $Ta_2NiSe_5$ at low temperature. Direct photoemission can be an important tool to characterize steady-state excitons.**

The concept of quasiparticles constitutes an essential building blocks of our understanding of solids, which allows us to unravel complex manybody interactions. An exciton, a bound state of an electron and a hole, is a popular quasiparticle in semiconductors, which is formed transiently as excited by light within the bandgap. This quasiparticle excitation not only let us understand the fundamental optical responses of semiconductors and photosynthesis[1], but also leads to various device applications[2], including photodetectors[3-6], photovoltaic solar cells[7-9], light emitters and excitonic lasers[10-13]. One can further make a macroscopically coherent condensate, that is, a Bose-Einstein condensate (BEC), of excitons using double layers containing electrons and holes separately in artificial structures with strong fields such as semiconductor quantum wells[14], graphene[15], and transition metal dichalcogenides[16,17]. The BEC of excitons may bring the above optoelectronic devices into an unprecedented coherent regime and is recognized to be important also in photosynthesis[1].

In stark contrast, there have been growing but debated experimental evidences[18-32] of excitons formed spontaneously without any external excitation or artificial structures to constitute an insulating ground state with a finite number of steady-state excitons[33-35]. Such an excitonic insulator (EI) is one fundamental manybody insulating state of solids, which was theoretically proposed as early as 1960's for a small gap semiconductor or a semimetal[33-35]. However, the accumulated evidences for the predicted properties of EI, such as diverging electronic susceptibility[32], anomalous optical transitions[36], quenching of plasmonic modes[18], and renormalized valence band dispersions[21], have been obscured by the coexisting structure phase transitions[37] and the direct experimental identification of spontaneously formed excitons has been missing.

Here we utilize temperature- and polarization (symmetry)-dependent angle-resolved photoemission spectroscopy (ARPES) to investigate the formation and the symmetry of the quasiparticles in a major EI candidate Ta$_2$NiSe$_5$ (TNS) across its anomalous phase transition. Our experimental results below the transition temperature ($T_c$ = 327 K) confirm the hybridization between the valence and the conduction band orbitals near the Fermi level. The origin of this interband hybridization has been debated with excitonic or structural origins. In contrast, the band structure above $T_c$ shows no clear sign of such band hybridization, ruling out a substantial structural effect, but the band gap persists. Moreover, we identify the emergence of a characteristic photoelectron signal, which can only be explained by the direct photoemission from excitons[38]. This is the first direct signature of the spontaneously formed excitons, the so-called 'preformed excitons' preceding the emergence of BEC at a lower temperature[39-44]. The present results, therefore, guarantee an excitonic insulator transition in Ta$_2$NiSe$_5$ at low temperature without any ambiguity related with the structural transition occurring at $T_c$ [39,44]. That is, this is in line with a recent theoretical proposal of a BEC-type EI transition for a semimetallic TNS[39] and indicates the existence of a novel manybody quantum state of preformed excitons at high temperature. Based on our ARPES result and a recent theory[38], the effective radius of excitons is extracted to be very small and anisotropic; ~7.5 Å and ~3.0 Å along parallel and perpendicular to Ta and Ni chains, respectively (Fig. 1a). This is attributed to the novel manybody interaction between excitons and phonons[36,45,46]. The present work opens up a new way to characterize spontaneously formed excitons with direct photoemission probes and to exploit coherent excitons at room temperature and preformed excitons at high temperature.

Figure 1b shows the overview of the band structure of TNS, revealed by our ARPES measurements along the $\bar{\Gamma}$-$\bar{X}$ and $\bar{\Gamma}$-$\bar{Y}$ directions (*x* and *y*-directions in Fig. 1a, respectively) over multiple Brillouin zones. The valence band near the Fermi level is characterized by a hole-like band centered at $\bar{\Gamma}$ with strongly anisotropic dispersions along the $\bar{\Gamma}$-$\bar{X}$ and $\bar{\Gamma}$-$\bar{Y}$ directions – a much stronger dispersion along the Ta/Ni atomic chains. As is well known, the band structure below $T_c$ is insulating with the valence band top located at about 0.16 eV (Fig. 1b). A distinct flattening of this valence band top, deviating significantly from a normal parabolic dispersion, is observed as shown in Fig. 2 in more detail. Together with the existence of the band gap, the band renormalization indicates a hybridization of the conduction and valence bands.

The interband hybridization itself can be more directly evidenced by revealing the orbital characteristic of the valence and conduction band edges, which are dominated by Ni 3*d* and Ta 5*d* orbitals, respectively, in the non-interacting case well above $T_c$[47]. This corresponds to the high-temperature orthorhombic crystal structure, where the hybridization between these two bands is forbidden by the crystal symmetry[47]. On the other hand, below $T_c$, the crystal structure changes to a monoclinic symmetry allowing the hybridization of Ni 3*d* and Ta 5*d* bands. The previous scanning tunneling spectroscopy study[24] indeed showed that the orbital (more accurately atomic) contributions of the valence (Ni 3*d*) and conduction band (Ta 5*d*) edges are inverted below $T_c$ [48]. This verifies the interband hybridization, which was interpreted as an indirect but important evidence of the EI state[21,22]. However, recent works pointed out that this hybridization itself can be explained largely by the orthorhombic-monoclinic structural transition[49].

ARPES with linearly polarized light can be utilized to determine the mirror symmetry of wave functions for occupied bands through the dipole selection rule in photoemission. We performed ARPES measurements along the $\bar{\Gamma}$-$\bar{X}$ direction for the low-temperature phase with four different linear polarization geometries (schematically shown in Fig. 1a). The data in Fig. 2a-d show that the photoelectron intensity of the valence bands exhibits a huge difference between the 'flat-top part' centered at the $\bar{\Gamma}$ point and the 'dispersive parts' away from $\bar{\Gamma}$, as summarized in Table 1. These apparent cross-sectional contrasts are independent of photon energies (50-90 eV, see Supplementary section 1) so that they are attributed to the symmetry selection rules between the light polarizations and orbital symmetries[50,51]. Based on the selection rules, the mirror parities of the valence and the conduction bands, respectively, are determined to be even (even) about *YZ*-plane and even (odd) about the *XZ*-plane (see Supplementary section 2 and 3). These observations are fully consistent with the hybridization of the valence and the conduction bands below $T_c$.

We then track the temperature dependence of the hybridization crossing the phase transition (see Supplementary section 4). Figure 3 shows the results of the polarization-dependent ARPES data above $T_c$ at 380 K. The flattening of the valence band top and the band gap persist well above $T_c$ (Figs. 3a,c). This is consistent with the earlier ARPES observations[23] and the resistivity measurements[45] indicating a finite band gap above $T_c$. However, in stark contrast to the low-temperature phase, the flat-top and the dispersive parts of the valence band exhibit the same polarization dependence; visible at *XZ*-polarization (Figs. 3a,c) and strongly suppressed at *Y*-polarization (Figs. 3b,d). This reveals unambiguously that, above $T_c$, the strong hybridization between the valence and the conduction band is significantly reduced for the flat-top of the valence band to retain largely the Ni 3*d* orbital character. This observation tells that the spontaneous orbital hybridization occurs only below $T_c$. Since $T_c$ was determined by the orthorhombic-monoclinic structural transition, the present result is consistent with the picture that the interband hybridization has a structural origin; the hybridization is indeed symmetrically forbidden in the orthorhombic structure[47]. On the other hand, since the non-interacting phase of TNS is semimetallic[39], there must be another interaction, which can explain the persistent band gap and the valence band renormalization (the flat top dispersion itself) above $T_c$.

In the following, we provide direct evidence of excitons formed spontaneously above $T_c$, which explain this extra interaction and most of the other anomalous behavior of this material. We focus on an extremely broad spectral feature at $\bar{\Gamma}$ on the flat-top part of the valence band (appearing clearly only above $T_c$ in the XZ-polarization) (see the dashed box in Fig. 3a). This feature is clearly distinguished from the dispersive parts away from $\bar{\Gamma}$, which is a usual spectroscopic feature for an electron band in ARPES. The broad feature show only marginal dispersion and does not cross the Fermi level (Figs. 4c,e). Moreover, as discussed above, this part does not have the symmetry of a Ta 5d orbital (as is clearly visible with XZ-polarization). Thus, the conduction band origin can be clearly ruled out in contrast to a previous suggestion[52] (see Supplementary section 2). The unusual broadening of this feature in both energy and momentum suggests an extraordinary origin beyond an electron band and substantial manybody interactions behind it.

Based on the discussion of the excitonic interaction in this material, we naturally check how a direct photoemission feature from excitons looks like. An excitonic feature in photoemission spectra was recently observed and characterized in detail with the pump-and-probe photoemission method as an excited state above the Fermi level and below the bottom of the conduction band[38,53-55]. We adopt this analysis established for pump-excited excitons while we are dealing with a spectroscopic feature below the Fermi level for a spontaneously formed steady-state object. We have performed the peak-fitting analyses using the two-dimensional excitonic wave functions $|\varphi(q)|^2 \sim 1/[1+(qa_0)^2/4]^4$ (where $q$ and $a_0$ are momentum and excitonic Bohr radius, respectively) for the momentum distribution curves (MDCs) at several energies across the broad feature at $\bar{\Gamma}$ (dashed lines in Fig. 4a)[38]. The results in Fig. 4b show excellent fits to all the observed MDCs across the entire binding energy range of 0.13-0.34 eV with a nearly constant excitonic Bohr radius of $a_0$ = 6.5-7.5 Å. Based on this excitonic Bohr radius we have also performed a model calculation for the ARPES spectra arising from the excitons based on the computational method presented by Rutsagi *et al.*[38] (with only one major phenomenological parameter of Lorentzian energy broadening of ~0.3 eV, which will be discussed further below). The result is shown in Fig. 4c for the excitonic contribution alone and in Fig. 4d together with the valence band dispersion (see Supplementary section 5). The overall shape of high-intensity region near the valence band maximum (i.e., resembling the bright up-side-down triangle) and the three-peak structure of MDCs in the ARPES data (Fig. 4e), can be reproduced excellently (Fig. 4f). These results convincingly indicate that we detect for the first time the direct photoemission signal from excitons formed spontaneously below the Fermi level.

The direct photoemission feature can make possible the detailed characterization of these excitons. We show in Figs. 5a and 5c the ARPES intensity plot along the $\bar{\Gamma} - \bar{Y}$ direction and the overview of the ARPES intensity plot in the momentum space near the Fermi level, respectively. These data reveal the giant anisotropy of the exciton photoemission feature as characterized by its much larger extent along the $\bar{\Gamma} - \bar{Y}$ direction than along $\bar{\Gamma} - \bar{X}$. The fitting analysis of MDC with an excitonic wave function, as shown in Fig. 5b, yields the excitonic Bohr radius along $\bar{\Gamma} - \bar{Y}$ to be about 3 Å being about only half of that along the other direction (Fig. 5d). This is comparable to the distance between the Ni and the Ta chain (~2.2 Å) and signals a strong localization of excitons perpendicular to the chains. The anisotropy is natural from the quasi one-dimensional structures in the lattice and the bands. This is also likely associated with strong and anisotropic coupling of excitons to phonons, by which the excitons can self-localize due to the short-range interactions, giving rise to the strong Fano resonance and the broadening of the absorption peak[36,46,56-58].

Another interesting observation is the significant broadening of the excitonic photoemission peak. The possible contributions for the broadening include (i) the intrinsic scattering of excitons, (ii) the broadening along $k_z$ arising from the finite extent of the excitonic wave function in the $z$-direction[53] and (iii) the finite experimental resolution. Since the experimental resolution is set to ~40 meV and the $k_z$-broadening is also a few tens of meV, limited by the dispersion width in the valence band in the Γ-Y direction (see Fig. 1b inset)[59], the sizable fraction of the observed width of ~0.3 eV at $\bar{\Gamma}$ (extracted from the $\bar{\Gamma}$ EDC in Fig. 3e) should be attributed to the intrinsic exciton scattering processes. Thus, the significantly broadened excitonic photoemission feature[22,23] is consistent with the presence of a strong exciton-phonon coupling. The anomalous broadening of the phonon modes identified in inelastic X-ray scattering and X-ray diffraction above $T_c$ [60] seems in line with the strong exciton-phonon coupling. The role of such a novel manybody interaction in the phase transition is a highly interesting topic of future research.

The existence of high temperature excitons above $T_c$ is in fact embedded as an essential ingredient of a BEC-type transition of excitons; excitons form at a much higher temperature than $T_c$ due to its strong interactions in a material with an EI instability and condense into a macroscopic quantum state at $T_c$. The 'preformed' excitons above $T_c$ are not coherent and the corresponding phase is a novel manybody quantum phase composed of spontaneously formed incoherent excitons. The present results thus evidence for the first time the preformed excitons and an emerging phase composed of them in a solid. In turn, it tells that the present material has a huge excitonic instability beyond the thermal and the structural energy scale and unambiguously guarantees that the Ta$_2$NiSe$_5$ represents an EI state at a sufficiently low temperature. We stress that this particular support of the EI cannot be associated with the structural transition occurring at $T_c$. The existence of preformed excitons is consistent with the very recent optical spectroscopy works[32] reporting the active electronic degree of freedom above $T_c$. On the other hand, the preformed excitons can explain naturally the persisting gap at high temperature, which was suggested as a counterevidence of the EI nature[49]. A BEC-like EI is in contrast to the other EI candidate of 1$T$-TiSe$_2$, for which a BCS-type excitonic instability is discussed[61]. This material exhibits a weak backfolding of the valence band above $T_c \sim 200$ K due to the effect electron-hole fluctuations[62], which is distinct from the direct exciton photoemission feature.

We finally note that the direct photoemission feature from excitons is not clearly observed in the EI phase below $T_c$. That is, in the low temperature EI phase, excitons do not seem to behave as individual quasiparticles as we assume for our analysis at higher temperature. This may be natural for a BEC in a macroscopic coherence[63,64], but no theory is available for the photoemission from a BEC of excitons at thermodynamic equilibrium at present. On the other hand, the strong interaction with the lattice (phonons) at low temperature may deviate the ground state well away from BEC as noted in a recent theoretical work[48] and the overall phase evolution follows the BEC behavior but in a heavily dressed way by the lattice degree of freedom[28].

In summary, we have utilized the symmetry-selective and temperature-dependent ARPES to identify the emergence of quasipaticles and their orbital symmetries near the Fermi level across the anomalous electronic transition in an EI candidate material of Ta$_2$NiSe$_5$. We identify unambiguously a characteristic photoemission feature coming directly from the excitons formed spontaneously above $T_c$ in the absence of the structural transition and in the presence of a persisting band gap. This clearly evidences the existence of the preformed excitons and fully validate the occurrence of an excitonic insulator phase transition in this material. Analyzing the direct excitonic photoemission feature reveals the compact and highly anisotropic size of excitons and strong scattering of them, which indicate the strong exciton-phonon coupling even in the precursor state of the excitonic insulator, signaling the importance of a novel higher-order manybody interaction in this material. By demonstrating the observability of steady-state excitons in ARPES, our study presents a new possibility for characterizing the excitonic properties not only in various types of excitonic insulator materials, but also for the future electronic and optoelectronic devices utilizing the presence of such steady-state quasiparticles.

# Methods

**Sample growth**

The single crystals of TNS were grown by chemical vapor transport method with iodine as a transport agent. The TNS polycrystalline sample, pre-reacted at 900 °C for 7 days, and iodine chunk (~2 mg/cm$^3$) are loaded in a quartz ampoule, which is placed in the two-zone furnace with a temperature gradient from 960°C to 850°C for 2 weeks. The as-grown single crystals are typically 2-3 x 0.1 x 0.01 mm$^3$, in a long bar shape along the *b*-axis, whose crystallinity and stoichiometry were confirmed by X-ray diffraction and energy-dispersive spectroscopy.

**Angle-resolved photoemission spectroscopy measurements**

The samples were prepared by *in-situ* cleaving to obtain a fresh surface parallel to the *xy*-plane, as shown in Fig. 1a. All ARPES measurements were performed at Pohang Light Source (PLS) BL-4A2. The photon polarizations of linear horizontal, linear vertical, right circular and left circular polarizations were utilized. Combining with the deflector acquisition modes, the samples were investigated with six different polarization geometries as schematically illustrated in Fig. 1a for linear polarizations and in Supplementary Fig. S2 for circular polarizations. For all of the measurements, the incidence angles of the light are the same. The photon energies of $h\nu$ = 50-90 eV were utilized and the temperatures of the measurements were set to $T$ = 75-400 K.

**Density functional theory calculations**

We use the experimental crystal structure of Ta$_2$NiSe$_5$ observed at room temperature and low temperature[60]. The unit cell contains eight Ta ions, four Ni ions and twenty Se ions. For relaxations of initial internal coordinates, the Vienna ab-initio Simulation Package (VASP), which employs the projector-augmented wave (PAW) basis set[66,67], was used for density functional theory (DFT) calculations in this work. The relaxation was performed on 14×3×4 Monkhorst-Pack k-mesh with a 400 eV energy cutoff. For the treatment of electron correlations within DFT, a revised Perdew-Burke-Ernzerhof exchange-correlation functional for crystalline solid (PBEsol) was employed[67], in addition augmented by on-site Coulomb interactions for transition metal *d*-orbitals within a simplified rotationally-invariant form of DFT+$U_{eff}$ formalism[68]. The correlation effect of the Ta and Ni *d* orbitals is investigated by observing the band dispersion and density of states (DOS) with various $U_{eff}$ values. The electronic structure of each phase (orthorhombic or monoclinic) varies widely from semi-metal to semiconductor with respect to varying $U_{eff}$ values. The $U_{eff}$ value of 2.25 eV for the Ta d orbitals and 4.75 eV for Ni d orbitals are adopted, which gives 27 meV semiconducting gap in monoclinic phase and 72 meV band overlap in orthorhombic phase. Structural relaxations were performed in the presence of the DFT+$U_{eff}$ on-site Coulomb interaction, all atomic coordinates were relaxed until forces exerted on all atoms are smaller than 0.01 eV/Å. Monoclinic phase is more stable than orthorhombic phase by 18 meV per unit cell.

**Acknowledgements:** The authors thank Byeong-Gyu Park for experimental helps and Changwon Park, Alexander Kemper and Avinash Rutsagi for fruitful discussions. This work was supported by the Institute for Basic Science (IBS), Korea, under project code IBS-R014-D01, and also by the National Research Foundation of Korea (NRF) through the SRC (No. 2018R1A5A6075964) and the Max Planck-POSTECH Center (No. 2016K1A4A4A01922028).

**Additional information:** Authors declare no competing financial interests.

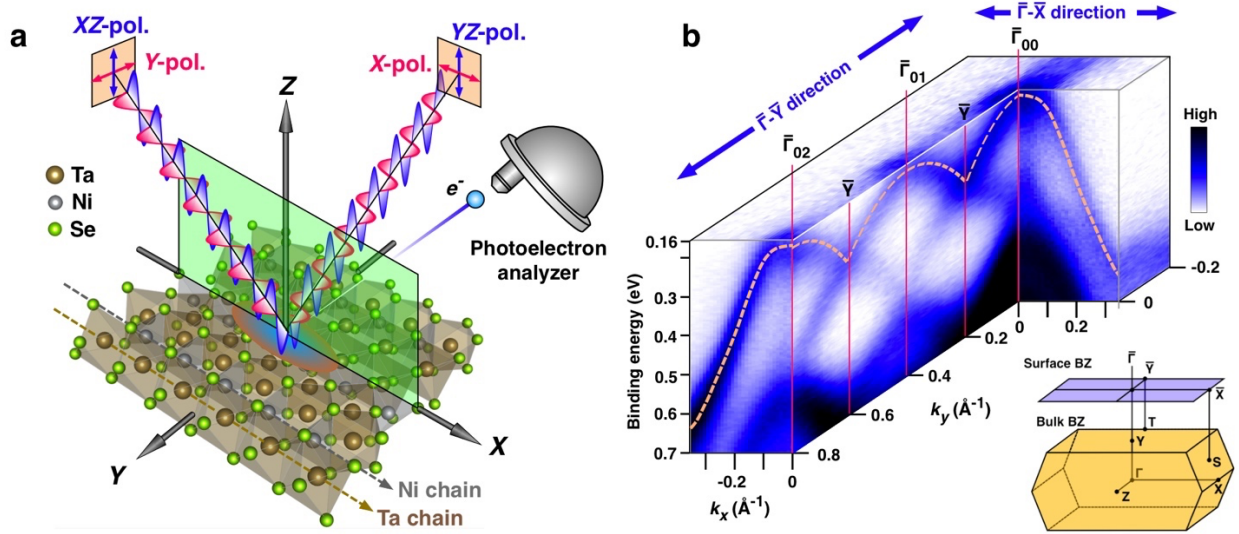

**Figure 1**: **The symmetry-selective ARPES experimental setup and band the structure of Ta$_2$NiSe$_5$. a,** Schematic illustrations of the experimental setup for the polarization-dependent (symmetry-selective) ARPES, where the four different linear polarizations (indicated as X-, Y-, XZ- and YZ-pol.) can be selectively utilized. The each of the polarizations are labelled based on the cartesian coordinate along which the electric field of the incident phonon is finite. The topmost layer of Ta$_2$NiSe$_5$ is also shown with the directions of Ta- and Ni atomic chains. The green plane indicates the plane of photoelectron detections along the $\bar{\Gamma}$-$\bar{X}$ direction used for the measurements shown in Fig. 2. **b,** The overview of electronic structure revealed by ARPES at the photon energy of $h\nu$ = 60 eV, $T$ = 75K taken with *YZ*-polarization as shown in **a**. The orange dashed curves are overlaid as the guide for the eyes to follow the dispersion of the topmost valence band. The inset at the lower right corner in **b** schematically illustrates the bulk (yellow) and surface (blue) Brillouin zones of Ta$_2$NiSe$_5$.

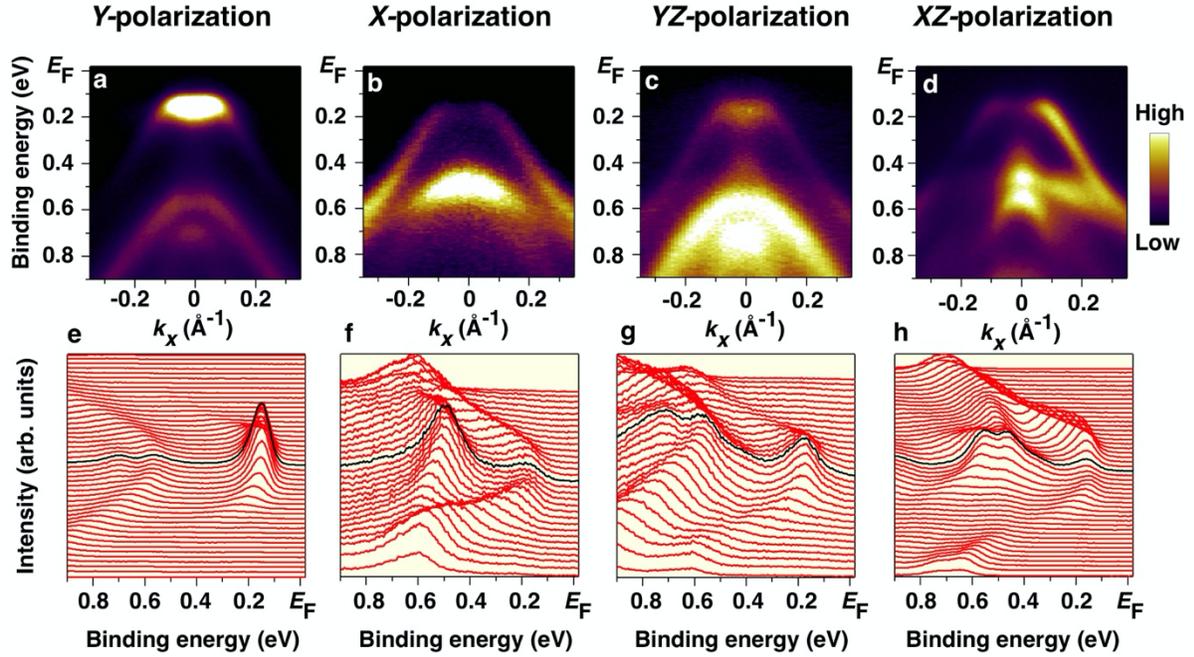

**Figure. 2: Results of the symmetry-selective ARPES.** The intensity plots of polarization-dependent ARPES results at the photon energy of $hv = 50$ eV and $T = 75$ K along the $\bar{\Gamma}$-$\bar{X}$ direction with the **a,** *Y*-polarization, **b,** *X*-polarization, **c,** *YZ*-polarization, **d.** *XZ*-polarization, as schematically illustrated in Fig. 1**a**. **e-h** show the corresponding stacked energy distribution curves (EDCs) extracted from the ARPES intensity plots in **a-d**, respectively.

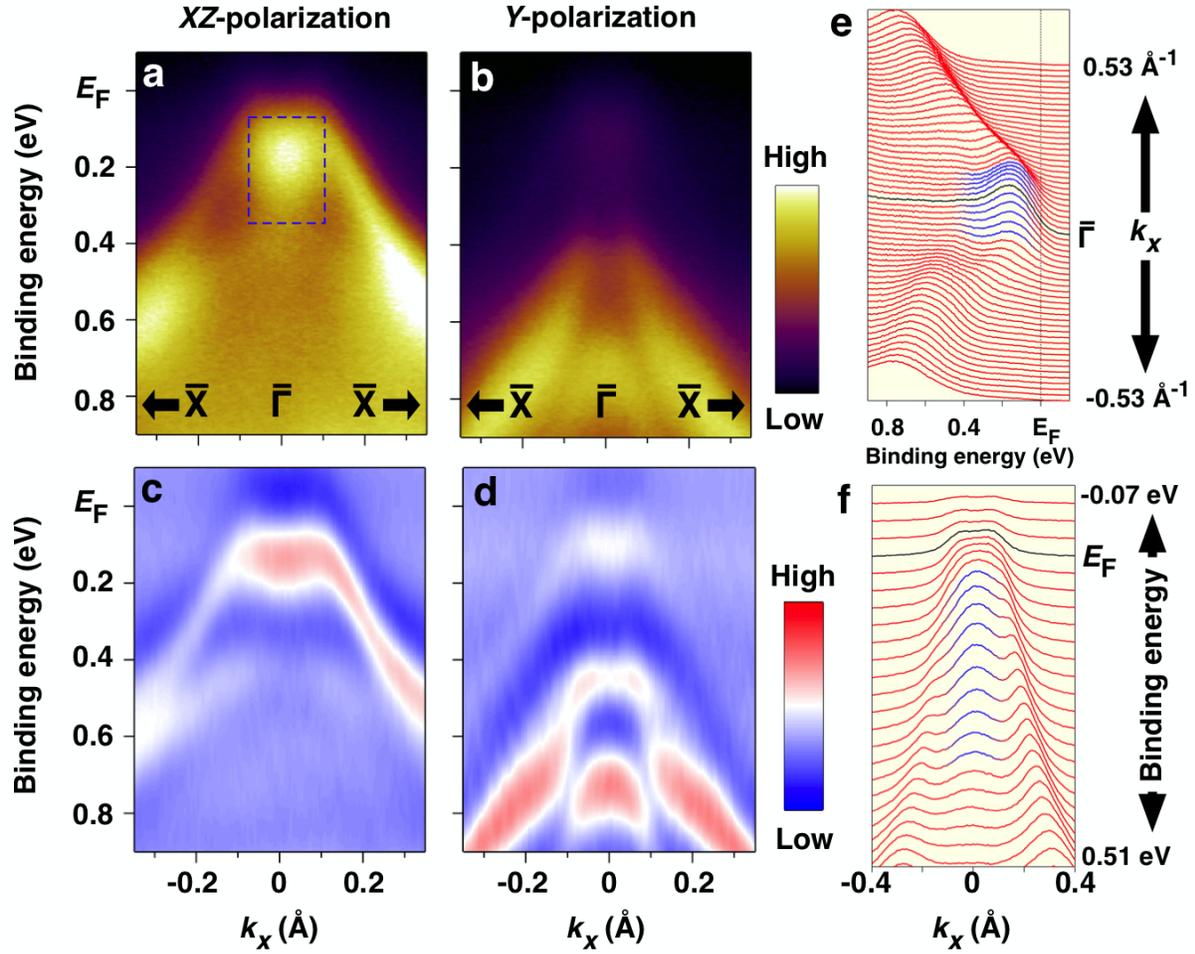

**Figure 3: Polarization-dependent ARPES results at $T$ = 380 K (above $T_c$).** **a** and **b** show the ARPES intensity plots for the measurements taken at $hv$ = 70 eV with the *XZ*- and *Y*-polarizations, respectively. The corresponding second derivative plots are shown in **c** for *XZ*-polarization and **d** for *Y*-polarization. **e** and **f** show, respectively, the stacked EDCs and MDCs for the further visualizations of intensity plot **a**. The characteristic strong signal around the $\bar{\Gamma}$ point is emphasized by the blue colors in the EDCs and MDCs in **e** and **f**, corresponding to the feature indicated by the dashed rectangle in **a**.

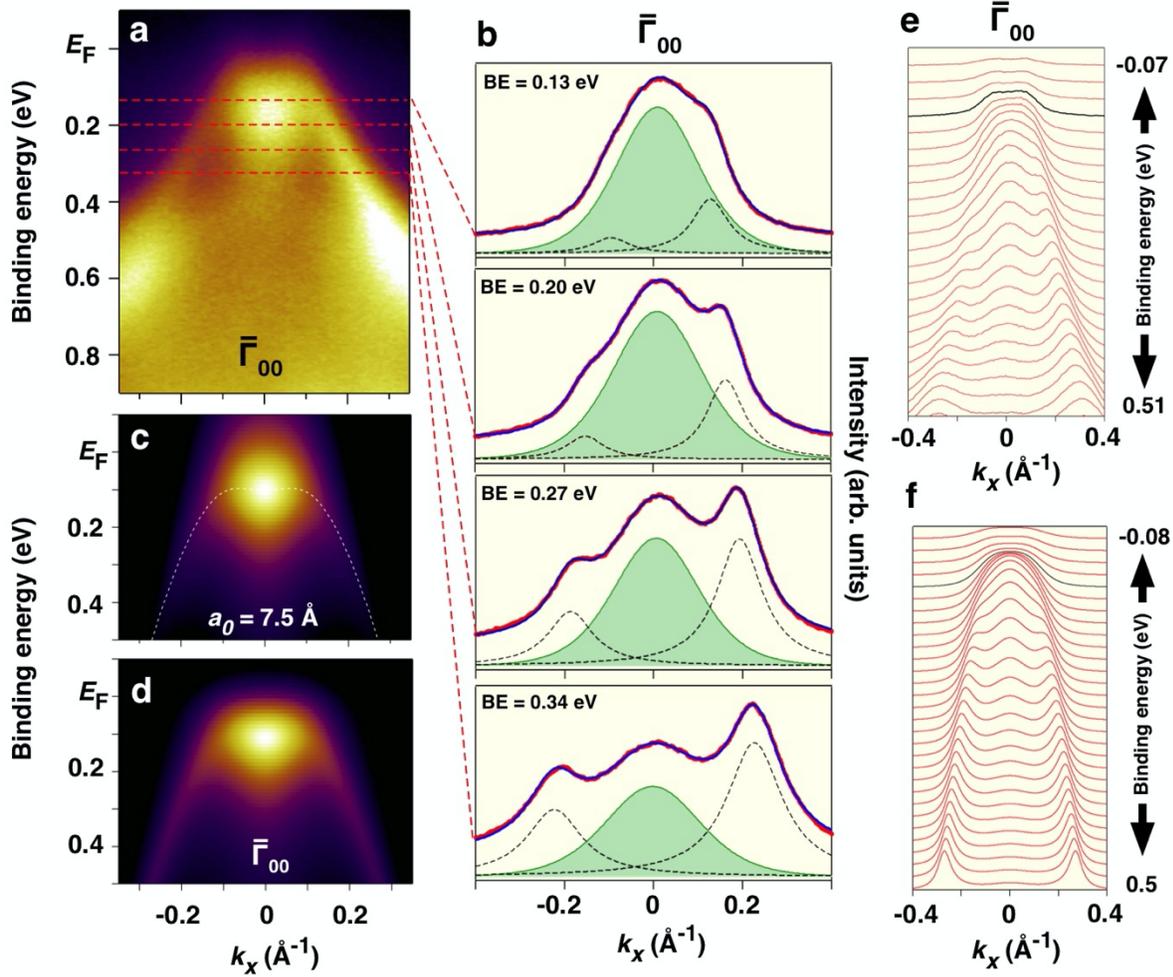

**Figure 4: The analysis of excitonic photoemissions at *T* = 380 K**. **a,** the ARPES intensity plot obtained at 380 K (same as in Fig. 3**a**), with the four dashed horizontal lines indicating the binding energies at which the MDC analysis shown in **b** was performed. **b** shows the MDC curve-fitting analysis performed at the binding energies of 0.13 eV, 0.20 eV, 0.27 eV and 0.34 eV indicated by the dashed horizontal lines in the intensity plot **a**. The green-shaded peaks centered at $\bar{\Gamma}$ show the best-fit excitonic wave function in momentum-space, mentioned in the main text, and the black dashed curves on both sides of $\bar{\Gamma}$ shows the Lorentzian fits to the valence band peaks (constrained to have the same width). The sum of the three peaks are shown as the blue curves which fit the red data points in each of the four panels. **c** shows the excitonic photoemission simulations based on the Bohr radius of 7.5 Å along the $\bar{\Gamma} - \bar{X}$ direction. **d** shows the sum of the valence band dispersion and the excitonic signal shown in **c** with the Fermi function. **e** and **f** show the stacked MDC curves for extracted from **a** and **d**, respectively.

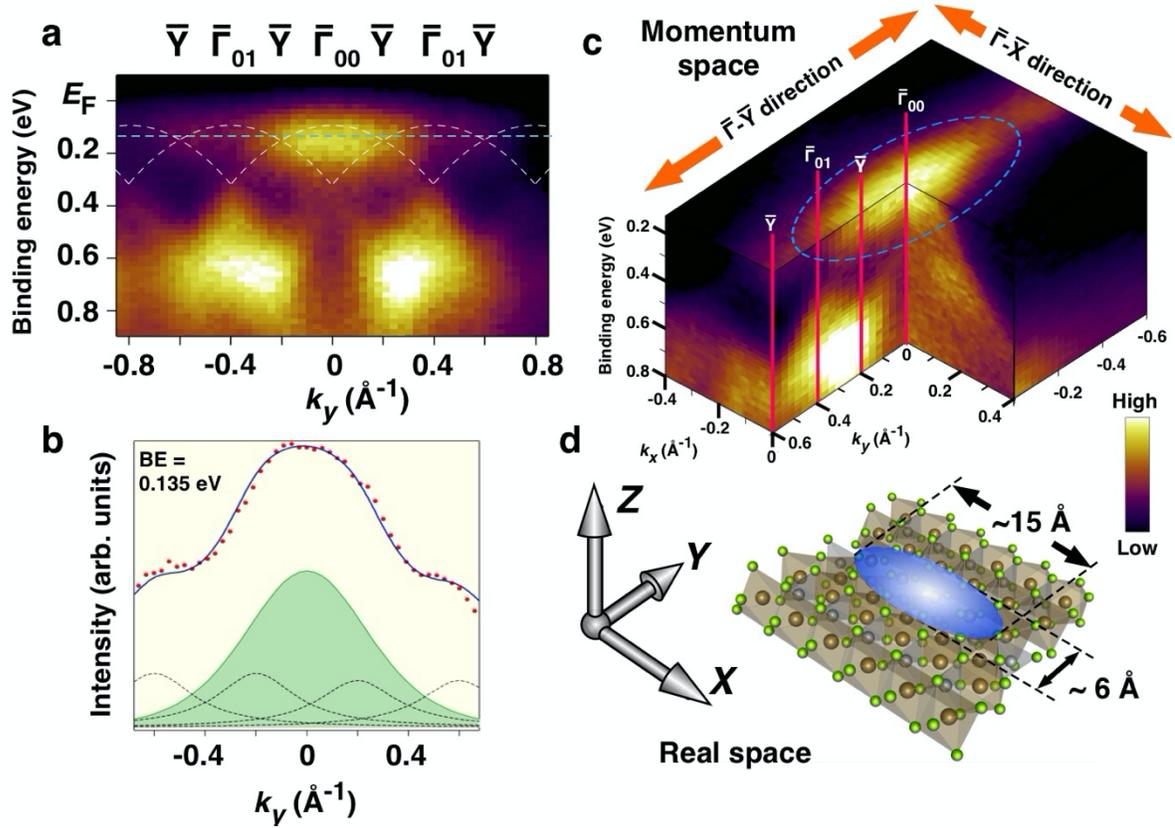

**Figure 5: The analysis of excitonic photoemission along $\bar{\Gamma} - \bar{Y}$ direction and the overview of preformed excitons in real and momentum space. a** shows the ARPES intensity plot along the $\bar{\Gamma} - \bar{Y}$ direction ($h\nu$ = 50 eV), where the guesstimates of the parabolic dispersions of the valence bands are overlaid as guides for the eyes. **b** shows the MDC curve-fitting analysis along the $\bar{\Gamma} - \bar{Y}$ direction (similarly performed as in $\bar{\Gamma} - \bar{X}$ direction) at the binding energy of 0.135 eV (the dashed horizontal line in **a**). **c** shows the $k_x$-$k_y$ map of the ARPES intensity as a function of binding energy ($h\nu$ = 50 eV), where the dashed ellipse indicates the significantly anisotropic photoemission feature arising from preformed excitons. The surface Brillouin zone centers are indicated as $\bar{\Gamma}_{00}$, $\bar{\Gamma}_{01}$. **d** shows the schematic drawing of the spatial extent of the exciton, extracted from the ARPES MDC analyses, scaled with respect to the underlying crystal structure.

# Supplementary Materials for

# Detecting photoelectrons from spontaneously formed excitons

Keisuke Fukutani[1,2], Roland Stania[1,2], Chang Il Kwon[1,3],
Jun Sung Kim[1,3], Ki Jeong Kong[1], Jaeyoung Kim[1,2], Han Woong Yeom[1,3*]

[1]Center for Artificial Low Dimensional Electronic Systems, Institute for Basic Science (IBS), Pohang 37673, Republic of Korea
[2]Pohang Accelerator Laboratory, Pohang University of Science and Technology (POSTECH), Pohang 37673 Republic of Korea
[3]Department of Physics, Pohang University of Science and Technology (POSTECH), Pohang 37673 Republic of Korea## 1. Photon energy independence of ARPES visibilities

Figure S1 shows the photon energy dependent ARPES intensity plots at $h\nu$ = 50-90 eV for $Y$- and $XZ$-polarizations. Apart from the variations of the relative intensities for different bands, naturally expected for different photon energies, the visibilities of the flat-top part and the dispersive parts of the valence band are unchanged over this photon energy range.

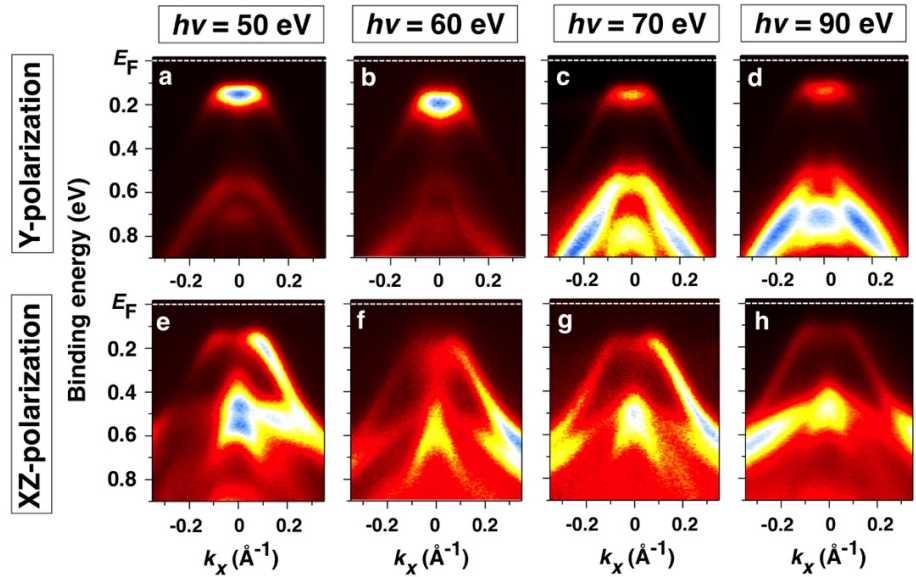

**Figure S1: The results of the photon-energy-dependent ARPES measurements for $Y$- and $XZ$-polarizations**. **a**-**d** show the ARPES intensity plots for $Y$-polarization at the photon energies of $h\nu$ = 50, 60, 70, 90 eV respectively. **e**-**h** show the ARPES intensity plots for $XZ$-polarization at the photon energies of $h\nu$ = 50, 60, 70, 90 eV respectively.

## 2. Identification of the mirror parities for the valence and conduction band orbitals

The photoemission spectra can be suppressed purely based on the symmetry considerations[S1,S2]. Here we show that the identified mirror parities for the valence band (VB) orbital and the conduction band (CB) orbital mentioned in the main text is the only combinations of the *xz*- and *yz*-mirror parities that are consistent with the band visibilities observed in our ARPES measurements.

The photoemission matrix element within the dipole approximation can be written as

$$\langle\psi_f|\boldsymbol{A}\cdot\boldsymbol{p}|\psi_0\rangle = \langle\psi_f|A_x p_x|\psi_0\rangle + \langle\psi_f|A_y p_y|\psi_0\rangle + \langle\psi_f|A_z p_z|\psi_0\rangle, \qquad (1)$$

where $\psi_i$ and $\psi_f$ are the initial and final states of electron, $\boldsymbol{A}$ is the vector potential of photon, and $\boldsymbol{p}$ is the momentum operator. The three terms on the right-hand side correspond to the photoemission matrix element arising from the *x*, *y*, and *z* components of the vector potential of photon (i.e., parallel to the photon polarization vector). Thus, for the photoemission matrix element to vanish, all of the three components must identically vanish. Each of the three terms in equation (1), can be rewritten in real space as

$$\langle\psi_f|A_i p_i|\psi_0\rangle = \iiint_{\mathfrak{R}^3} \psi_f(x,y,z)[A_i p_i]\psi_0(x,y,z)\,dxdydz, \qquad (2)$$

where *i* = *x*, *y*, or *z*. Thus, the vanishing condition for the photoemission matrix element in (2) is that at least one of the spatial integral (with respect to *x*, *y*, or *z*) vanishes due to parity considerations (i.e., an integrand is odd). The results of this parity consideration and whether the photoemission matrix element must vanish or not for the four possible combinations of the mirror parities for the initial state are summarized in the Tables S1-S4 below. In the following tables, the *yz*- and *xz*-parities of the initial and final states of electrons are expressed as | *yz*-parity, *xz*-parity >, where "+", "-", and "x" indicate even, odd, and no (i.e., neither even nor odd) mirror parities, respectively.

From the photoemission selections rules summarized in Table S1-S4 above, it can be seen that there is no single pair of *yz*- and *xz*-parities, consistent with the observed visibilities in ARPES at $T<T_c$ for the flat-top part at $\bar{\Gamma}$ and the dispersive parts away from $\bar{\Gamma}$ in $\bar{\Gamma}$-$\bar{X}$ direction (Fig. 2). Thus, we identify the pair of mirror parities at $\bar{\Gamma}$ and $\bar{\Gamma}$-$\bar{X}$ line separately. Since the flat-top part is clearly visible only for *Y*- and *YZ*-polarizations (i.e., the polarization containing *Y*-component), it can be deduced that the dominant orbital at $\bar{\Gamma}$ (that of conduction band orbital) should have the $|+->$ parities. For the dispersive parts away from $\bar{\Gamma}$ (predominantly of valence band orbital), they are clearly visible in *XZ*-, *YZ*-, and *X*-polarizations, but not in *Y*-polarization. Thus, it can be deduced that the *xz*-parity must be even (see Table S1 and S2). The *yz*-parity for the valence band orbital can be deduced based on the observation that the photoemission intensity at the $\bar{\Gamma}$ point (flat-top part) is weakly visible at *XZ*-polarization, but nearly vanishing in *X*-polarization. Since the valence band orbital contribution should be weaker than the conduction band orbital (due to band inversion), but not vanishing at $\bar{\Gamma}$, it can be deduced that *Z*-component of light should be responsible for the weak visibility of the valence band at $\bar{\Gamma}$ in the *XZ*-polarization. Thus, the *yz*-parity of the valence band orbital should be identified as even; that is, the set of parity of valence band orbital at $\bar{\Gamma}$ should be identified as $|++>$ (see Table S2).

Here, we note that the mirror parities about *XZ*- and *YZ*-plane are not strictly good quantum numbers in the monoclinic structure ($T < T_c$) but the selection rules based on the orthorhombic symmetry well explain the observed ARPES polarization dependence. We believe that the monoclinic distortion is too small (less than 1°)[S3] to affect considerably the photoemission cross section of wave functions.

As a further verification that the difference in the band visibilities for the "flat-top" part and the dispersive parts are unique to the linearly polarized incident lights, which selectively probes the wave functions with the compatible mirror symmetries, we have also performed the ARPES measurements with circularly polarized light. Since the photoemission selection rules based on the mirror parities does not give any general restrictions for the circular polarizations, we expect that both flat-top and the dispersive parts should be visible. Such expectations are indeed confirmed as shown in Fig. S2 shown below.

**Table S1.** The photoemission selection rule for | *yz*-parity, *xz*-parity > = |- +> at $\bar{\Gamma}$

| Photon polarization | Matrix element at $\bar{\Gamma}$ | Matrix element between $\bar{\Gamma}$ and $\bar{X}$ |
|---|---|---|
| X | $\langle + + \vert - + \vert - + \rangle \neq 0$ | $\langle \times + \vert - + \vert \times + \rangle \neq 0$ |
| Y | $\langle + + \vert + - \vert - + \rangle = 0$ | $\langle \times + \vert + - \vert \times + \rangle = 0$ |
| Z | $\langle + + \vert + + \vert - + \rangle = 0$ | $\langle \times + \vert + + \vert \times + \rangle \neq 0$ |

**Table S2.** The photoemission selection rule for | *yz*-parity, *xz*-parity > = |++> at $\bar{\Gamma}$

| Photon polarization | Matrix element at $\bar{\Gamma}$ | Matrix element between $\bar{\Gamma}$ and $\bar{X}$ |
|---|---|---|
| X | $\langle + + \vert - + \vert + + \rangle = 0$ | $\langle \times + \vert - + \vert \times + \rangle \neq 0$ |
| Y | $\langle + + \vert + - \vert + + \rangle = 0$ | $\langle \times + \vert + - \vert \times + \rangle = 0$ |
| Z | $\langle + + \vert + + \vert + + \rangle \neq 0$ | $\langle \times + \vert + + \vert \times + \rangle \neq 0$ |

**Table S3.** The photoemission selection rule for | *yz*-parity, *xz*-parity > = |+ -> at $\bar{\Gamma}$

| Photon polarization | Matrix element at $\bar{\Gamma}$ | Matrix element between $\bar{\Gamma}$ and $\bar{X}$ |
|---|---|---|
| X | $\langle + + \vert - + \vert + - \rangle = 0$ | $\langle \times + \vert - + \vert \times - \rangle = 0$ |
| Y | $\langle + + \vert + - \vert + - \rangle \neq 0$ | $\langle \times + \vert + - \vert \times - \rangle \neq 0$ |
| Z | $\langle + + \vert + + \vert + - \rangle = 0$ | $\langle \times + \vert + + \vert \times - \rangle = 0$ |

**Table S4.** The photoemission selection rule for | *yz*-parity, *xz*-parity > = |- -> at $\bar{\Gamma}$

| Photon polarization | Matrix element at $\bar{\Gamma}$ | Matrix element between $\bar{\Gamma}$ and $\bar{X}$ |
|---|---|---|
| X | $\langle + + \vert - + \vert - - \rangle = 0$ | $\langle \times + \vert - + \vert \times - \rangle = 0$ |
| Y | $\langle + + \vert + - \vert - - \rangle = 0$ | $\langle \times + \vert + - \vert \times - \rangle \neq 0$ |
| Z | $\langle + + \vert + + \vert - - \rangle = 0$ | $\langle \times + \vert + + \vert \times - \rangle = 0$ |

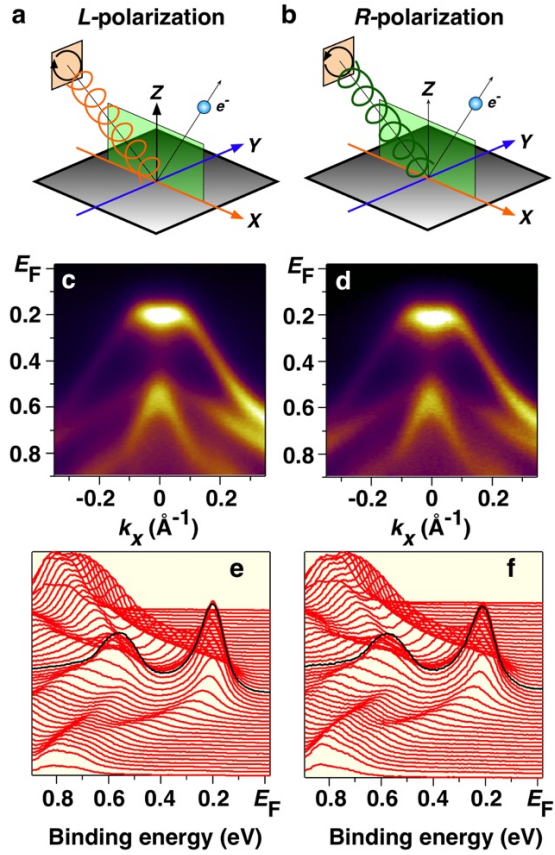

**Figure S2**: **The ARPES results for taken for $\bar{\varGamma} - \bar{X}$ line with the left (L) and right (R) circular polarizations.** **a** and **b** show the schematics of the measurement geometries. **c** and **d** show the ARPES intensity plots. **e** and **f** show the corresponding EDCs for L- and R-polarizations respectively.

## 3. Symmetries of the valence and the conduction bands from DFT calculations

In order to support the experimental identifications of the symmetries of the valence and conduction band orbitals, we have performed the *ab-initio* band structure calculation for $Ta_2NiSe_5$ in the monoclinic phase using density functional theory. Figure S3a shows the calculated band structure near the Fermi level in the monoclinic phase. To infer the mirror parities ($\sigma_{yz}$, $\sigma_{xz}$) of the electronic states for the conduction and the valence band with respect to the *yz*- and the *xz*-planes, the dominant *d*-orbital components of the wave functions were extracted at the Γ point for the conduction band, and the two points (flat-top part and the dispersive part) for the valence band, as indicated by red, green and orange circles in Fig. S3a. As shown in Figs. S3b and S3c, the wave function at the bottom of the conduction band is dominantly composed of $d_{x2-y2}$ and $d_{yz}$ orbitals for Ni and Ta atoms, and it can be seen that the wave function composed of these orbitals have the parities of ($\sigma_{yz}$, $\sigma_{xz}$) = (even, even). Similarly, as shown in Figs. S3d and S3e, the wave function at the strongly hybridized flat-top part of the valence band is dominantly composed of $d_{z2}$ and $d_{x2-y2}$ orbitals for Ni and Ta atoms, and it can be seen that the wave function composed of these orbitals must have the parities of ($\sigma_{yz}$, $\sigma_{xz}$) = (even, odd), consistent with our ARPES identification. Furthermore, as shown in Figs. S3f-h, the wave function at the dispersive part of the valence band is dominantly composed of $d_{z2}$, $d_{xz}$, and $d_{xy}$ orbitals, and it can be seen that the $\sigma_{xz}$ parity is even. The results that the flat-top part exhibiting the opposite $\sigma_{xz}$ parity from the dispersive part but the same parity as the conduction band is consistent with the picture of strong hybridization (inversion) between the valence and conduction bands, and is consistent with our symmetry identifications based on ARPES.

We note that while these symmetry identifications are different from the ones presented[S4,S5], the point of view that the valence and the conduction band possess different set of mirror symmetries are common to these and our studies. Our experimental and theoretical identifications, in the mutual agreement, correctly assign the mirror parities of the conduction and the valence bands.

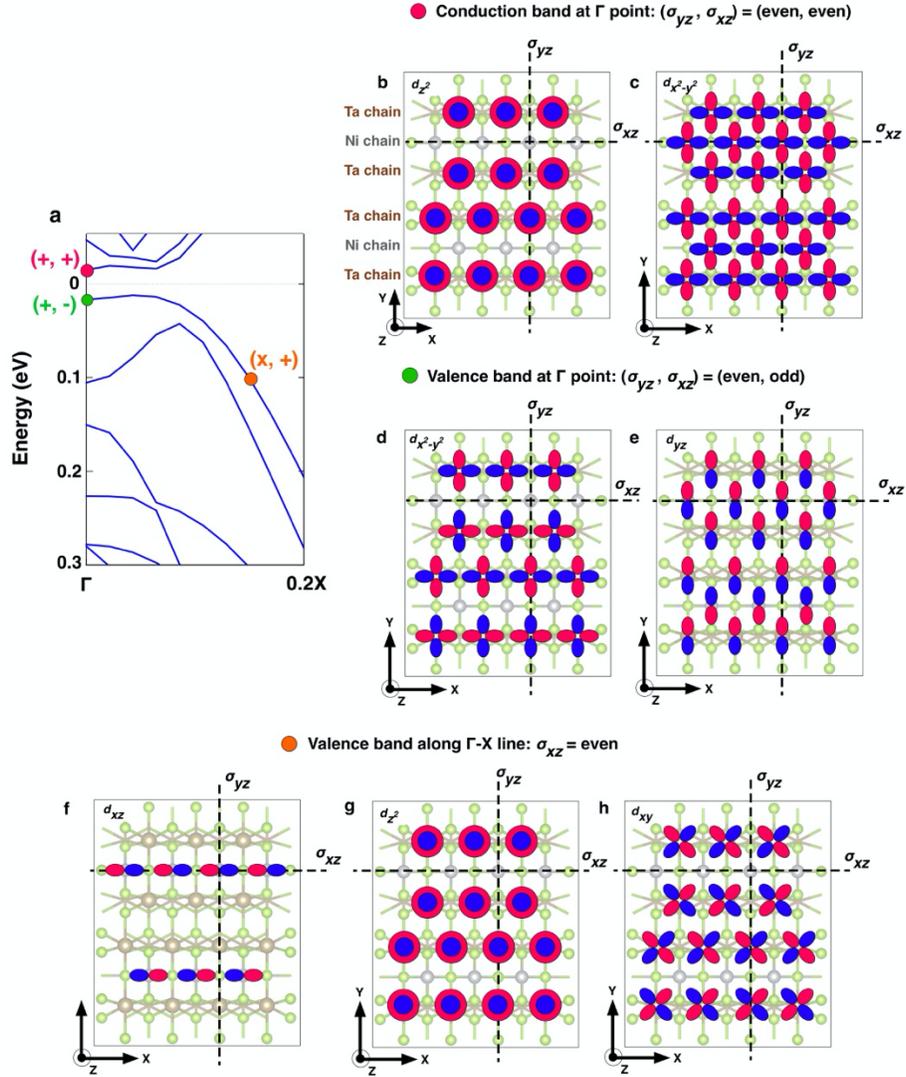

**Figure S3**: **Calculated band structure of Ta$_2$NiSe$_5$ and the dominant *d*-orbitals. a** The calculated band structure of Ta$_2$NiSe$_5$ in the monoclinic phase along the Γ-X direction. The inferred mirror parities ($\sigma_{yz}$, $\sigma_{xz}$) of the conduction band at Γ (red circle), valence band (green circle at Γ and orange circle along the Γ-X line) are indicated. For each of these points the schematic illustrations of the predominant orbital components are shown for (**b-c**) conduction band, (**d-e**) the flat-top part and (**f-h**) the dispersive part of the valence band. The red and blue colors in (**b-h**) represent opposite wave function phases.

## 4. Temperature-dependent ARPES results across the transition temperature

Figure S4 below shows the ARPES intensity plots taken at the temperature of $T$ = 280-400 K. It can be clearly seen that the intensity plots for the temperature below $T_c$ (Figs. S4a and S4b) show the strong suppression of the intensity near top of the valence band at the $\bar{\Gamma}$ point (due to the photoemission selection rules mentioned in the main text and Section 2 above). This is qualitatively same as the observation at $T$ = 70 K. In contrast, the intensity plots for the temperatures above $T_c$ (Fig. S4c-e) clearly show the appearance of the strong intensity at the $\bar{\Gamma}$ point near the top of the valence band.

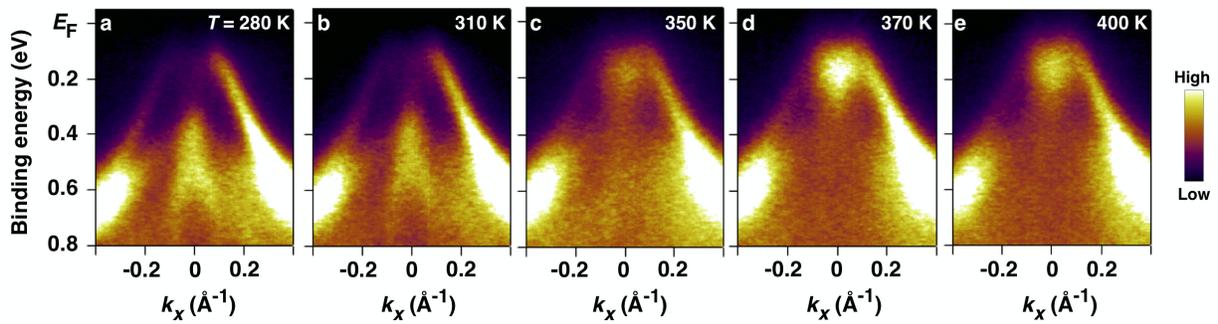

**Figure S4: Temperature dependence of the ARPES intensity plots**. **a-e** show ARPES intensity plots at the indicated temperatures at $T$ = 280, 310, 350, 370 and 400 K, respectively, across the phase transition temperature $T_c$ = 327 K, taken at $hv$ = 70 eV with *XZ*-polarization.

## 5. Model calculations for the photoemission signals from excitons

In order to simulate how the photoemission intensity should appear when there are finite number of exciton present in the material (preformed exciton phase), we have followed the theoretical model calculation scheme presented in Rustagi *et al.*[S6], designed for time-resolved photoemission after the excitons have been created by the pump pulse. The model equation is [S6]:

$$P \cong 2\pi\sigma^2 |M_{k,k'}|^2 \sum_{\lambda,Q} \rho_{\lambda,Q} |\phi_{\lambda,w}(k' - w - \alpha Q)|^2 \exp\left(-\sigma^2[-\omega + E_{\lambda,Q} + \epsilon_{v,k'-w-Q}]^2\right) \quad (1),$$

where the parameters are defined as follows:

$P$ = Photoemission intensity
$\sigma$ = Temporal width of the probe pulse
$M_{k,k'}$ = Photoemission matrix element (assumed to be constant, for simplicity)
$\rho_{\lambda,Q}$ = Exciton distribution function
$\phi_{\lambda,w}$ = Exciton wave function in momentum space $\propto 1/[1+p^2 a_0^2/4]^2$, where $a_0$ is the exciton Bohr radius and $p$ is momentum.
$\omega$ = Energy referenced to the valence band maximum
$E_{\lambda,Q}$ = Exciton band dispersion
$\varepsilon_v$ = Valence band dispersion
$Q$ = Exciton center-of-mass momentum
$\alpha = m_c/(m_c + m_v)$, where $m_c$ and $m_v$ are the conduction band mass and valence band mass in the effective band mass approximation
$w$ = Momentum separation between the valence band maximum and conduction band minimum (= 0 for direct gap material TNS)

In order to perform the computations for TNS above the EI transition temperature, several assumptions have been made. First, since the exciton band is theoretically predicted to cross the valence band maximum at $T_c$ to induce the excitonic instability[S7], the exciton band should lie close to the valence band maximum. For this reason, we have placed the exciton band so that its bottom coincides with the valence band maximum in TNS. For the exciton distribution function, since the BEC-type EI transition is characterized by condensation of zero-momentum excitons formed above $T_c$ [S8], we only compute the photoemission signature arising from the excitons with $Q = 0$. For this reason, the knowledge of the exciton band dispersion (at finite $Q$), which is not exactly known for TNS becomes unnecessary for the purpose of our computations. The equation (1) takes into account the effect of finite temporal width of the pulse in the Gaussian shape (the exponential factor at the end of the equation), which gives rise to the finite linewidth in the computed ARPES spectra. On the other hand, in our measurement, the pulse width (train bunch width at the synchrotron) is estimated to be ~ 20 ps and this gives rise to nearly completely negligible contribution from the temporal-width-induced broadening. On the other hand, in solids, the exciton can be scattered by phonons, impurities etc. and such process can induce the finite broadening, which will be reflected in the finite width of the actual ARPES spectra. Thus, the exponential factor at the end of equation (1) has been replaced by the experimentally determined Lorentzian function with constant width of 0.3 eV. The photoemission matrix element $M_{k,k'}$ was taken to be constant (independent of momentum) for the computations. Based on these assumptions and the equation (1) above, the spectral functions for photoemission from the excitons have been computed.